\RequirePackage[l2tabu, orthodox]{nag}
\documentclass[twocolumn,amsmath,amssymb,aps,showkeys,nofootinbib]{revtex4-2}

\usepackage{graphicx}
\graphicspath{{figures/}}
\usepackage{dcolumn}
\usepackage{bm}
\usepackage[hidelinks]{hyperref}

\usepackage{titlesec}
\titleformat{\section}[hang]{\normalfont\bfseries}{\thesection.}{0.5em}{}[]
\titlespacing{\section}{0pt}{*2}{*1}

\begin{document}

\title{Seasonality and susceptibility from measles time series}
\author{Niket Thakkar}
\email{niket.thakkar@gatesfoundation.org}
\homepage[\\]{https://github.com/NThakkar-IDM/seasonality}
\author{Sonia Jindal}
\author{Katherine Rosenfeld}
\affiliation{%
The Institute for Disease Modeling | Bill $\&$ Melinda Gates Foundation\\
Seattle, Washington 98109
}%
\date{\today}

\begin{abstract}
This paper develops mathematical tools to estimate seasonal changes in measles transmission rates and corresponding variation in population susceptibility. The tools are designed to leverage times series of cases in the absence of demographic data. In particular, we focus on publicly available suspected case reports from the World Health Organization (WHO), which routinely publishes country-level, monthly aggregated time series. With that as input, we show that measles epidemiologies can be characterized efficiently at global-scale, and we use our estimates to recommend context-specific, future supplementary immunization times. Throughout the paper, comparisons with more data-informed models illustrate that the approach captures the essential dynamics, and broadly speaking, the tools we describe represent a scalable intermediate between conventional empirical approaches and more intricate disease models. 
\end{abstract}
\keywords{disease modeling, measles, vaccination, inference, forecasting, stability analysis}
\maketitle

\section{Introduction}

The observed shapes of measles epidemics -- that rises tend to begin in the winter and that they're often multi-peaked -- was the initial inspiration to look for seasonal variation in transmission rates \cite{soper1929interpretation,london1973recurrent}. Researchers closely inspected time series and demographic data from New York and Baltimore, from England and Wales, and from Glasgow, and they eventually uncovered durable, annually periodic dynamics that were evident both before and after vaccine introduction \cite{fine1982measles}. 

The mechanism driving these dynamics was debated. Some researchers hypothesized for example that the measles virus itself was changing periodically \cite{fine1979john}. But over time, with particular attention paid to the time-scales of the disease, the details of the seasonality profiles in these settings reinforced existing ideas that transmission was modulated by school terms and holidays \cite{fine1982measles}. 

More generally across contexts the drivers of seasonal variation depend on societal structure, but the idea that stable ecological features shape the measles seasonality profile continues to be foundational to our understanding of measles transmission processes \cite{finkenstadt2000time,ferrari2008dynamics,bharti2011explaining,thakkar2019decreasing,fu2021effect}. Uncovering underlying mechanisms bespoke to certain settings remains an important part of public health, helping guide vaccine delivery to the right populations at the right times.

Despite this history, measles seasonality profile estimation tends to be one component of broader transmission modeling studies, and public health officials more often rely on histograms of case reports over time to distinguish between high- and low-transmission seasons \cite{cliff1980changes,li2012epidemic,omonijo2012effect}. Approachable methods, ones applicable to standardized data that retain the simplicity of empirical tools while capturing the essence of transmission models, are needed to bridge this gap between science and practice.

This paper addresses this inference problem. Specifically, we develop a regression method to estimate seasonality profiles from publicly-available, monthly-aggregated time series of suspected cases \cite{whodata}. The method approximately accounts for the 2 week measles generation time and the realities of underreporting \cite{finkenstadt2000time,thakkar2022modeling}, and we show through comparisons with more involved models that the inferred profiles capture the essential dynamics.

Then, taking our estimate as representative of a setting's recent epidemiology, we show that the seasonality profile has a corresponding stable, endemic transmission dynamic, one that can be studied to construct context-specific estimates of susceptibility variation. Tracking both susceptibility and seasonality gives us coarse estimates of outbreak start times, which we show perform well on some historical examples. Taken together, these two statistical estimators applied to time series data can be used to determine critical times for vaccine delivery. We discuss those implications in detail.

\section{Transmission seasonality}

\begin{figure}
\centering\includegraphics[width=\linewidth]{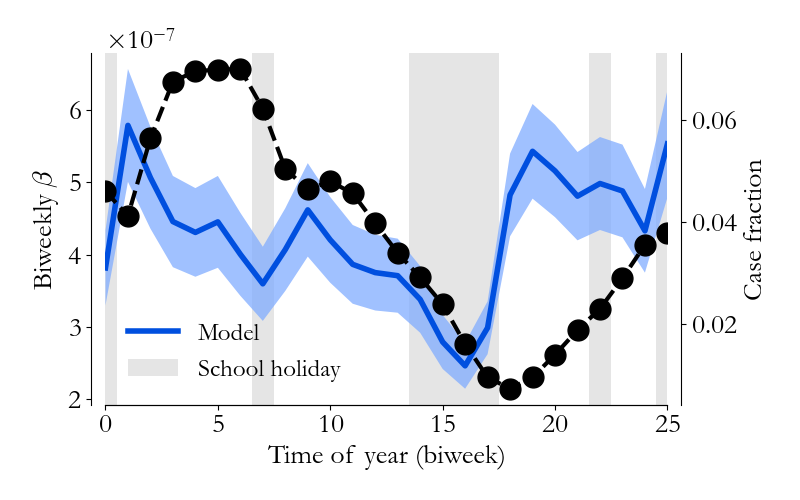}
\caption{The seasonality profile estimate (blue) for England and Wales, reproduced from Ref. \onlinecite{finkenstadt2000time}, illustrates that measles case histograms (black) can miss key features like the association with schools (grey).}
\label{fig:uk}
\end{figure}

The history mentioned above as well as some context to get us started are encapsulated in Fig. \ref{fig:uk}, a modified version of Ref. \onlinecite{finkenstadt2000time}'s Fig. 7. In blue is the famous estimate of the measles seasonality profile in England and Wales. Transmission rates have pronounced declines when school is out of session (grey) followed by rebounds when students return. Meanwhile, the corresponding histogram of cases is overlaid in black. With only that statistic, the association with schools vanishes and estimates of the high- and low-seasons are lagged. 

Keeping Fig. \ref{fig:uk} in mind, our approach in this section is to first define the seasonality profile in terms of a transmission model and then to work our way towards the monthly aggregated time series we intend to study. Following Refs. \onlinecite{finkenstadt2000time,bharti2011explaining,ferrari2008dynamics,thakkar2019decreasing}, consider a susceptible population, $S_t$, and an infectious population, $I_t$, interacting in a stochastic measles transmission process on semi-monthly increments, $t$. If we approximate the generation time of measles deterministically as one semi-month, susceptible-infectious interactions lead to new infections as
\begin{equation}
    I_{t+1} = \beta_t S_t I_t, \label{eq:sir}
\end{equation}
where the transmission rate, $\beta_t$, is assumed to be log normally distributed with an annually periodic mean which we call the seasonality profile.\footnote{In many applications, including Refs. \onlinecite{finkenstadt2000time,bharti2011explaining,ferrari2008dynamics,thakkar2019decreasing}, Eq. \ref{eq:sir} includes an extra exponent, $\beta_t S_t I_t^{\alpha}$, with $\alpha \leq 1$. This exponent models the idea that some infectious individuals cannot contribute to onward infection -- a conventional example is a younger sibling of a child who just started school. In sensitivity testing, we find $\alpha$ is needed for stable extrapolation, but less important for inference. Here we choose to set $\alpha = 1$ with that in mind.} Then, the corresponding susceptible population evolves according to
\begin{equation}
    S_{t+1} = S_t + B_t - I_{t+1} - V_t, \label{eq:St}
\end{equation}
where $B_t$ represents new susceptible individuals being born, and $V_t$ accounts for immunization through vaccination campaigns, to be discussed in more detail later. Finally, we model observed cases, $C_t$, as
\begin{equation}
    C_t \sim \text{Binom}\left\{I_t, p_t\right\}, \label{eq:binom}
\end{equation}
with unknown and in principle time-varying detection probability $p_t$. Then if $t=0$ marks the end of a month, what we observe is a monthly time series $C_{2m-1} + C_{2m}$ for months $m=0, 1, 2, ..., T$.

The corresponding infectious population dynamics can be constructed through repeated compositions of Eq. \ref{eq:sir} and then simplified by approximating $S_t \approx S_{t+1}$ as slowly varying, giving
\begin{equation*}
    I_{t+2} + I_{t+1} \approx (\beta_t S_t)^2(I_{t} + I_{t-1})\left[1 + \mathcal{O}\left(\frac{1}{I_{t} + I_{t-1}}\right)\right].
\end{equation*}
For country-scale applications, with generally large populations, we can safely neglect the error term, and we're left with an intuitive, first-order result that monthly groups of infectious individuals approximately interact with the same susceptible population at the same rate twice. From here on, we take that approximation, and we use $t$ to refer to months.

To isolate the seasonality profile, we write $S_t = \hat{S} + Z_t$, considering contexts where susceptibility has small fluctuations, $Z_t$, around a large endemic average, $\hat{S}$, so that $\ln(S_t)\approx \ln(\hat{S}) + Z_t/\hat{S}$. Then, along similar lines, we note from Eq. \ref{eq:binom} that $I_t$ is approximately Gaussian in the large population limit, with average $\hat{I}_t = (C_t + 1)/p_t - 1$ and variance that scales as $1/(C_t + 1)$ relative to the mean \cite{thakkar2022modeling}. When outbreaks are happening at this scale, $C_t$ is large, and we can approximate $\ln I_t \approx \ln\hat{I}_t$.

With these approximations in hand, using $\ln\beta_t = \ln\hat{\beta}_t + \varepsilon_t$ with mean $\ln\hat{\beta}_t$ and volatility $\varepsilon_t$, we have
\begin{equation*}
    \ln(\hat{\beta}_t\hat{S}) + \varepsilon_t + \frac{Z_t}{\hat{S}}
    \approx \frac{1}{2}\ln\left(\frac{p_{t}}{p_{t+1}}\frac{C_{t+1}+1-p_{t+1}}{C_{t}+1-p_{t}}\right).
\end{equation*}
Finally, with the goal in mind to analyze timeseries data alone, we approximate $p_t \approx p_{t+1}$, that is that the reporting rate varies slowly month-to-month. Further, we assume reporting is low, so $1-p_t \approx 1$, and we assume transmission volatility $\varepsilon_t$ has constant variance that dominates the contribution from $Z_t/\hat{S}$. We're left with
\begin{equation}
    \ln R_t + \varepsilon_t \approx \frac{1}{2}\ln\frac{C_{t+1}+1}{C_t + 1}, \label{eq:regression}
\end{equation}
where we've defined $\hat{\beta}_t\hat{S} \equiv R_t$ as the effective reproductive number. Eq. \ref{eq:regression} is a tractable Gaussian process regression for seasonal variation in $R_t$ given a monthly aggregated timeseries $C_t$. We can solve this regression problem with standard methods \cite{bishopbook,sivia2006data}, incorporating periodic month-to-month correlation in $R_t$.

\begin{figure*}
\centering\includegraphics[width=\linewidth]{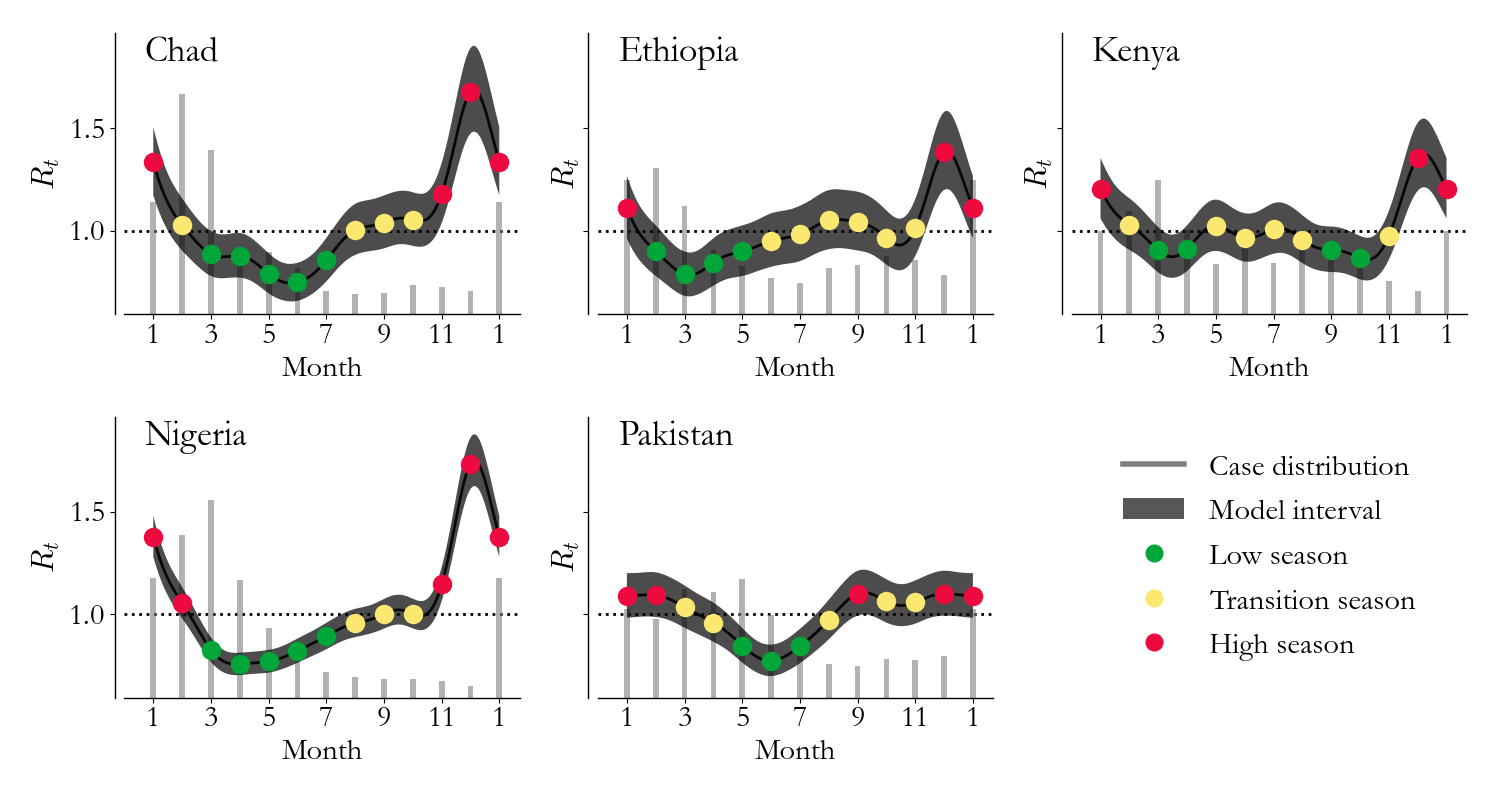}
\caption{Seasonality profiles (dark grey curve) estimated via Eq. \ref{eq:regression} offer more nuanced and quantitative perspective on low- (green) and high-season (red) timing than corresponding histograms of cases (grey bars). Variation in seasonality across settings points to the context-specific nature of the underlying mechanisms.}
\label{fig:profiles}
\end{figure*}

Application of this approach to data from a few countries is visualized in Fig. \ref{fig:profiles}. Seasonality profiles are associated with month-to-month rises and falls in the histograms (grey bars), but with faster variation and a shift in phase as we saw in Fig. \ref{fig:uk}. Across countries, profiles take on a variety of shapes, but with some shared features like peaks from November to January. 

\begin{figure}
\centering\includegraphics[width=\linewidth]{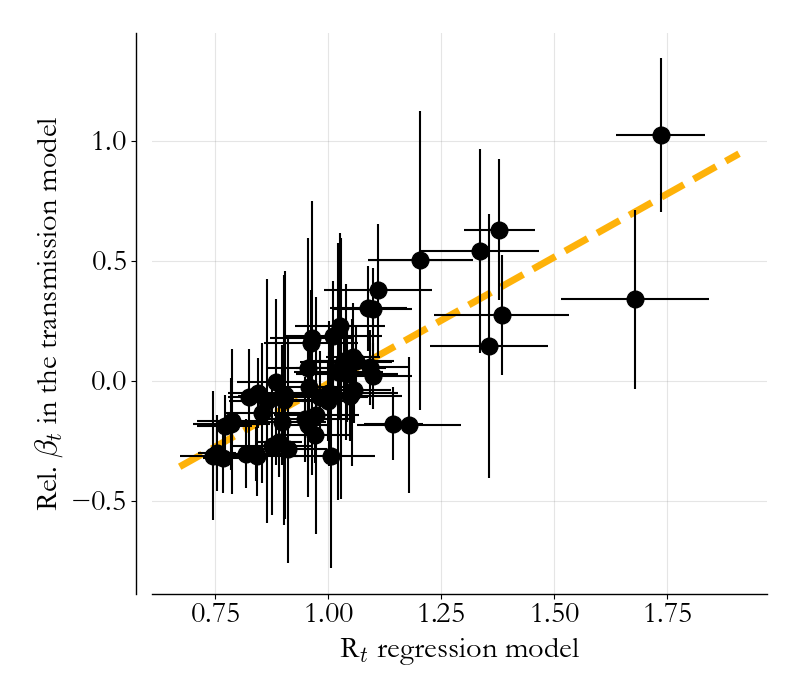}
\caption{Comparing the estimates from Fig. \ref{fig:profiles} to those from more data-informed, mechanistic models of measles transmission shows reasonable overall agreement.}
\label{fig:vs_tsir}
\end{figure}

Finally, marked as dots, we can use Eq. \ref{eq:regression} to estimate the probability that $R_t < 1$ in a given month, giving us a quantitative measure of low- and high-seasons. In Fig. \ref{fig:profiles}, we choose $\text{Pr}(R_t < 1) \geq 0.8$ as the definition of low-season (green), $1 - \text{Pr}(R_t < 1) \geq 0.8$ as the definition of high-season (red), and anything else as a transition season (yellow). Across settings, low and high seasons are shorter and earlier than what might be estimated from the histogram. 

We can compare the estimates in Fig. \ref{fig:profiles} to those from a more informed transmission model to give us some confidence in our approximations. We use the model from Ref. \onlinecite{thakkar2019decreasing} adapted to include independent campaign effectiveness parameters for each vaccination campaign in the study period. The model is applied at the semi-monthly timescale, fit to interpolated $C_t$, leveraging demographic and routine immunization coverage estimates from the World Bank \cite{wbdata} and the WHO \cite{wuenicdata}. 

Fig. \ref{fig:vs_tsir} compares the seasonality profile estimates from Fig. \ref{fig:profiles} to those from the more intricate model. The high degree of correlation (Pearson coefficient of $0.80$) across the full range of $R_t$ values is reassuring. While we made some coarse assumptions to arrive at Eq. \ref{eq:regression}, we appear to have achieved reasonable consistency.

\section{Endemic stability}

Seasonality profiles give us empirical estimates of the times of year when outbreaks tend to begin, but a well-studied feature of measles epidemiology is that despite annually periodic transmission rates, outbreaks have multi-year periodicities \cite{fine1982measles}. Guided by transmission models, researchers have shown that the growth rate of the susceptible population, $B_t$ above, sets this slower dynamic timescale \cite{finkenstadt2000time,ferrari2008dynamics}. Intuitively, outbreaks are most likely to occur when both transmission rates and susceptibility are high.

Estimating $B_t$, particularly after vaccine introduction, is generally the role of demography and coverage survey in epidemiological studies \cite{thakkar2019decreasing,fu2021effect}. But the seasonality profile gives us a complementary perspective. Starting with Eq. \ref{eq:St}, we can solve the recursion relation to find
\begin{equation}
    S_t = S_0 + \sum_{i = 0}^{t-1}\left( B_t - V_t \right) - \sum_{i = 1}^{t} I_t, \label{eq:St_solved}
\end{equation}
which in our monthly approximation to Eq. \ref{eq:sir} implies
\begin{equation*}
    I_{t+1} \approx R_t^2\left(1 + \frac{S_0-\hat{S}}{\hat{S}} + \mathbf{A}_0 \frac{(B_t-V_t)}{\hat{S}} - \mathbf{A}_1 \frac{(I_t)}{\hat{S}}\right)^2 I_t,
\end{equation*}
where we've implicitly defined the linear operator $\mathbf{A}_i (\cdot) \equiv \sum_{j = i}^{t-1+i} (\cdot)$. In the endemic equilibrium, $\hat{S}$ is the largest quantity, and if we neglect terms $\mathcal{O}(1/\hat{S})$, we find the endemic equilibrium incidence
\begin{equation}
    I_t = I_0 \prod_{i=0}^{t-1} R_i^2,\label{eq:It_endemic}
\end{equation}
which tells us that the infectious population grows and declines with the seasonality profile as we would expect. Inserting this result into Eq. \ref{eq:St_solved}, setting $V_t = 0$ in keeping with the equilibrium assumptions, and rearranging gives
\begin{equation*}
    S_t - \hat{S} = S_0 - \hat{S} + \mathbf{A}_0 B_t - \mathbf{A}_1 I_0 \prod_{i=0}^{t-1} R_i^2,
\end{equation*}
for the corresponding deviations from average susceptibility. Then, noting that the left-hand-side has 0 average, we can take the expected value of both sides and rearrange to find
\begin{equation*}
    \frac{S_0 - \hat{S}}{I_0} +  \mathbf{A}_0 \frac{\hat{B}_t}{I_0} = \mathbf{A}_1 \prod_{i=0}^{t-1} R_i^2,
\end{equation*}
which intuitively states that growth in susceptibility needs to balance seasonal incidence to maintain equilibrium. If we assume $\hat{B}_t = B$, that is a constant average growth rate, we have
\begin{equation}
    \alpha_0 +  \alpha_1 t = \mathbf{A}_1 \prod_{i=0}^{t-1} R_i^2, \label{eq:stability}
\end{equation}
where we've defined $\alpha_0 \equiv (S_0 - \hat{S})/I_0$ and $\alpha_1 \equiv B/I_0$ to clarify that demographic factors are related to accumulated linear trend in the seasonality profile, up to an unknown but constant factor $I_0$. 

\begin{figure}
\centering\includegraphics[width=\linewidth]{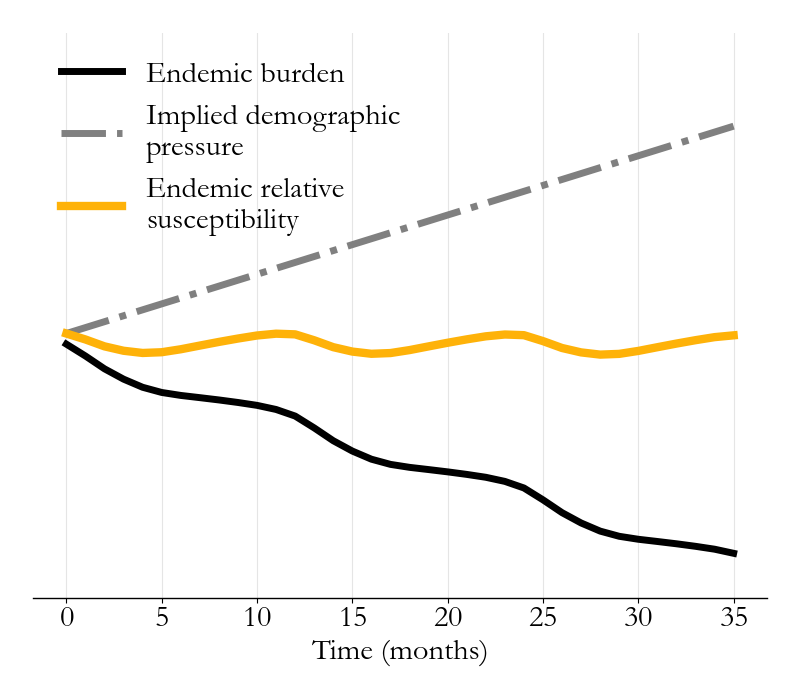}
\caption{Seasonality profiles dictate the shape of the long-time, average endemic process (yellow), giving us a window into balance between susceptible population growth (grey) and infection (black).}
\label{fig:stability}
\end{figure}

The logic of Eq. \ref{eq:stability} is illustrated in Fig. \ref{fig:stability}. Endemic burden (black) must be periodic with variation due to seasonality in order to maintain equilibrium and consistency with a setting's profile. But simultaneously, demographic pressure (grey) represented by $\alpha_1$ must balance burden to maintain a stable susceptible population (yellow). In other words, Eq. \ref{eq:stability} gives us a statistic of the seasonality profile that approximates the usual demographic inputs to transmission processes. 

This motivates another comparison to the more detailed transmission models. Specifically, extrapolating a fitted model to its endemic equilibrium, we can calculate the average variation in susceptibility and deviation from that average across sample trajectories. Eqs. \ref{eq:regression} and Eq. \ref{eq:stability} should capture the same dynamics. 

\begin{figure*}
\centering\includegraphics[width=\linewidth]{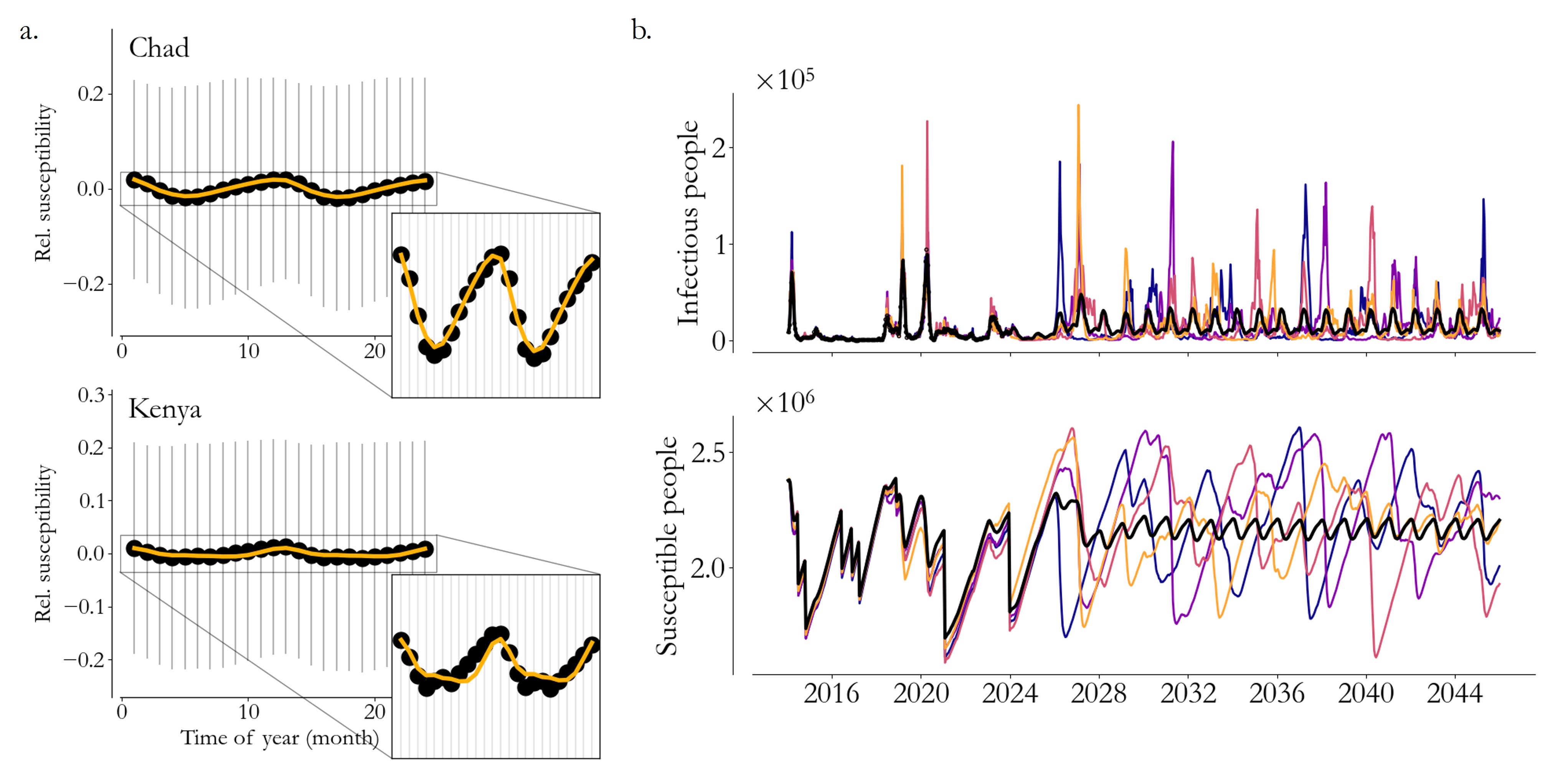}
\caption{(a) Profile-based estimates (yellow) capture the average endemic fluctuations in susceptibility (black dots) from the more complex transmission models, but variance around that average (2 standard deviation error bars) is the more significant consideration. (b) Sample trajectories from the transmission model for Chad show that the long-time average (black) is never actually realized. But still, the black curves' existence gives us practical estimates of susceptible growth rates.}
\label{fig:vs_endemic}
\end{figure*}

This comparison is visualized for Chad and Kenya in Fig. \ref{fig:vs_endemic}a. Black dots are the average variation around $\hat{S}$ across 10000 trajectories, with error bars visualizing 2 standard deviations around that average. Meanwhile, estimates based on the seasonality profiles from Fig. \ref{fig:profiles} are scaled and overlaid in yellow. Two features stand out: the stability argument underlying Eq. \ref{eq:stability} clearly captures the average dynamic, but volatility around that average dominates variation in the black dots.

The situation is clarified with a closer look at the Chad transmission model in Fig. \ref{fig:vs_endemic}b. In black is the average $I_t$ and $S_t$ across trajectories, fit to data up to 2024 and then extrapolated to the endemic equilibrium by roughly 2030. In color are 5 individual trajectories from the model. While the data constrains trajectories up to 2024, no trajectory follows the mean in the future. Individual outbreaks throw the process off the average, leading to rebounds and recoveries that separate large outbreaks for years. So while we can explain the long-time behavior of the black curves, they have limited direct utility in terms of context-specific risk assessment and epidemiology.

\section{Susceptibility reconstruction}

We shouldn't be discouraged by Fig. \ref{fig:vs_endemic}b. While it's certainly the case that the endemic average is never realized in practice, its existence teaches us practical details about the transmission process. We can revisit the data with those in mind.

More specifically, returning to Eq. \ref{eq:St_solved} with $\alpha_0$ and $\alpha_1$ in hand, we can write
\begin{equation*}
    \frac{S_t-\hat{S}}{I_0} = \alpha_0 + \alpha_1 t - \frac{1}{I_0}\mathbf{A}_1 I_t - \frac{1}{I_0} \mathbf{A}_0 V_t,
\end{equation*}
and we can ask for tractable models of $I_t$ and $V_t$, ones that maintain simplicity along the lines of Eqs. \ref{eq:regression} and \ref{eq:stability}. 

For $V_t$, the WHO publishes records of immunisation campaigns with estimates of the number of doses delivered, $d_t$. We can coarsely approximate $V_t \approx \mu d_t$, where $\mu$ is an unknown, time-averaged immunizing fraction of those doses. Meanwhile, inspired by Ref. \onlinecite{finkenstadt2000time}, we can choose $I_t \approx \rho C_t$, with some unknown, constant scale-factor $\rho$ corresponding to the average $1/p_t$ from Eq. \ref{eq:binom}. Inserting these models into the equation above and rearranging gives
\begin{equation}
    \frac{\rho}{I_0}\mathbf{A}_1 C_t + \frac{\mu}{I_0} \mathbf{A}_0 d_t + \frac{S_t-\hat{S}}{I_0} = \alpha_0 + \alpha_1 t, \label{eq:reg2}
\end{equation}
a linear regression for $\rho/I_0$ and $\mu/I_0$, noting that $S_t-\hat{S}/I_0$ is a mean-zero volatility. 

Once solved, the residuals from the regression above are very nearly an interpretable measure of susceptibility fluctuations, $Z_t/\hat{S}$ in the discussion of Eq. \ref{eq:regression}. But unfortunately the dependence on $I_0$ obfuscates the scale. To convert to more meaningful units, we can return to the monthly transmission model, and note that
\begin{equation*}
    I_{t+1} = R_t^2\left(1 + \frac{I_0}{\hat{S}}\frac{S_t-\hat{S}}{I_0}\right)^2 I_t.
\end{equation*}
Taking the log, rearranging, and then exponentiating again gives
\begin{equation*}
    \frac{I_0}{\hat{S}}\frac{Z_t}{I_0} = \exp\left(\frac{1}{2}\ln\frac{I_{t+1}}{I_t} - \ln R_t \right) - 1,
\end{equation*}
which approximately relates the residuals from Eq. \ref{eq:regression} to those from Eq. \ref{eq:reg2}. In other words, we can compute $Z_t/I_0$ directly, and then rescale to $Z_t/\hat{S}$, fluctuations as a fraction of the average susceptible population, by projecting onto the residuals around the seasonality profile.

\begin{figure*}
\centering\includegraphics[width=\linewidth]{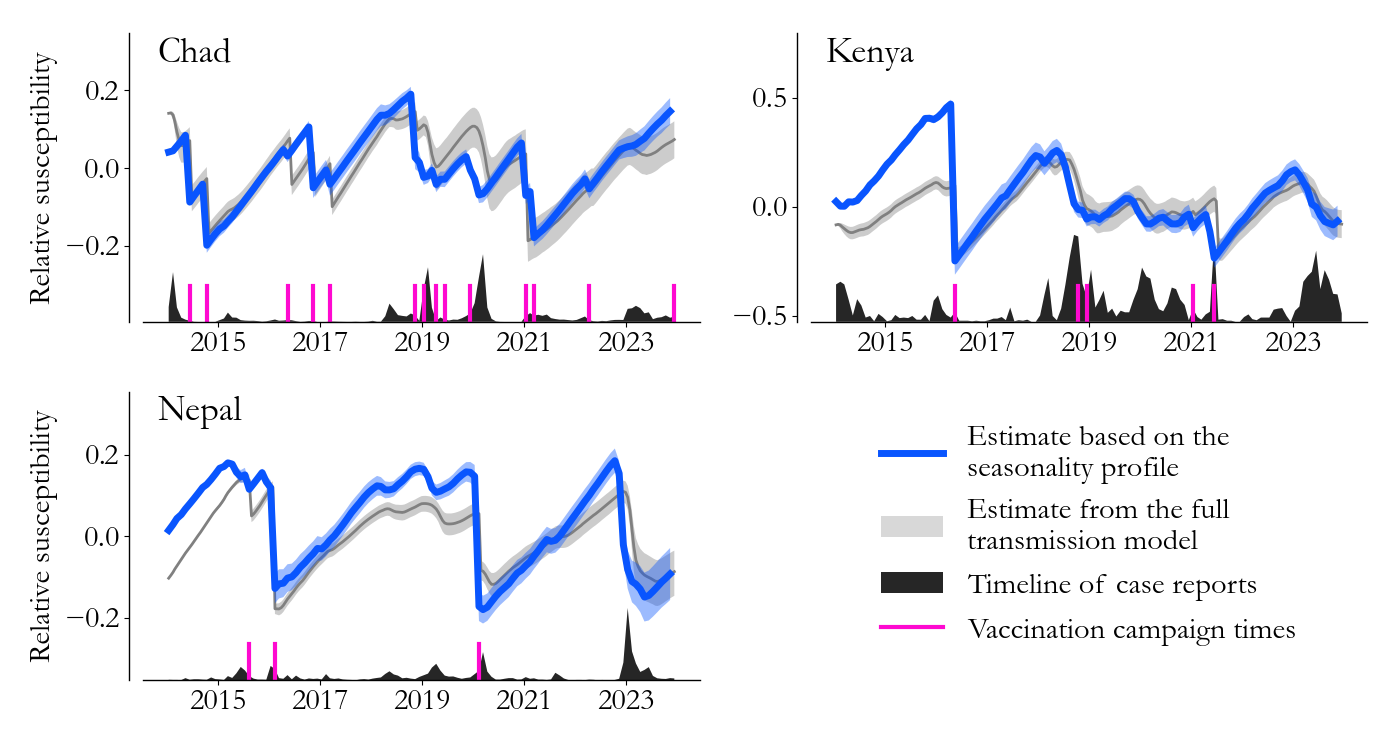}
\caption{Relative susceptibility estimates for Chad, Kenya, and Nepal based on their respective seasonality profiles (blue) have reasonable agreement with estimates from the more informed transmission models (grey). The effects of vaccination campaigns (pink) and large outbreaks (black) on susceptibility are apparent, and growth in susceptibility estimated via Eq. \ref{eq:stability} agrees well with the demographic estimates incorporated into the transmission models.}
\label{fig:sus_recon}
\end{figure*}

\begin{figure*}
\centering\includegraphics[width=\linewidth]{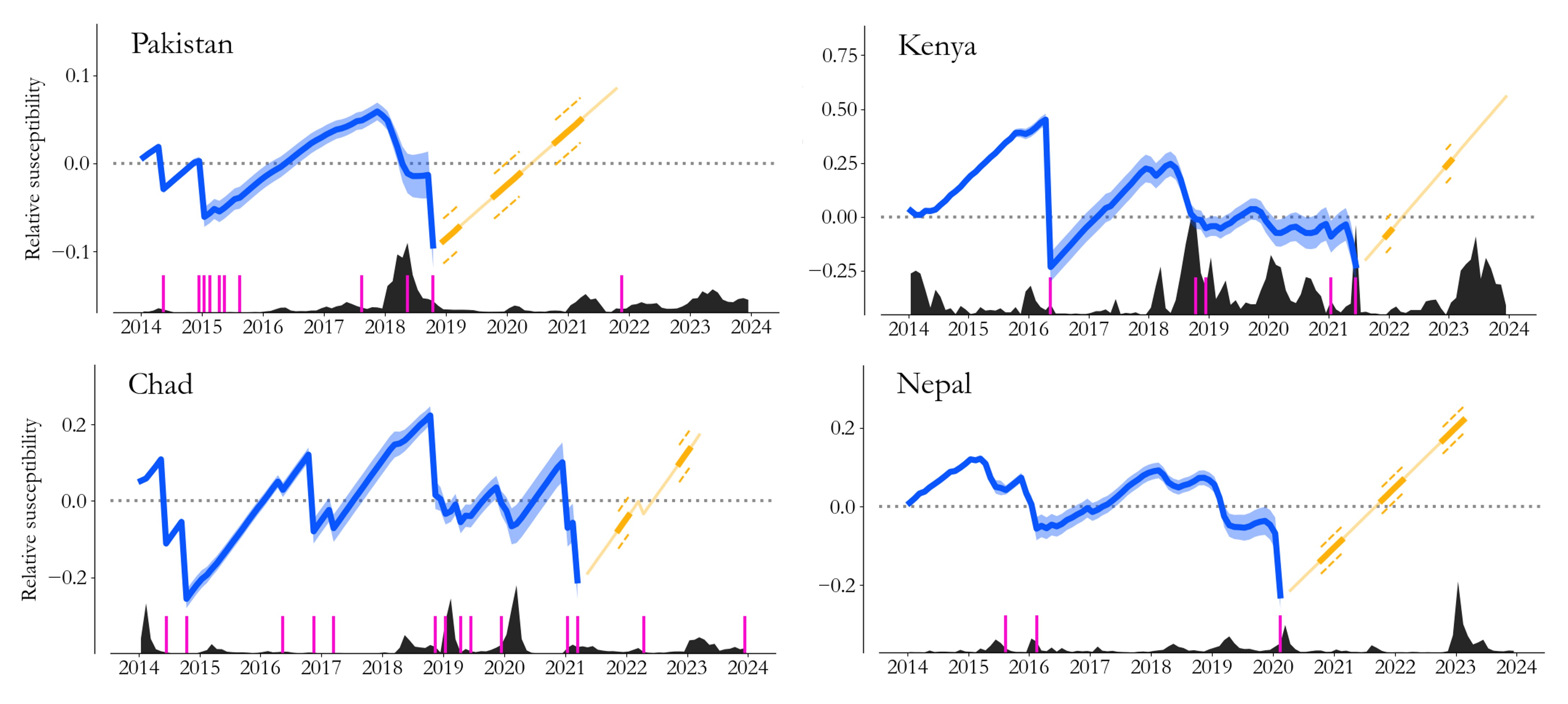}
\caption{Hypothetical examples from Pakistan, Kenya, Chad, and Nepal show that relative susceptibility (blue) can be calculated in the wake of a large event like a campaign or outbreak. The corresponding estimates of susceptibility growth (orange) can be used to approximate the future time till above average $S_t$ (grey dotted line). Combined with the seasonality profile estimate, we can use this coarse linear forecasting approach to anticipate high-susceptibility high-transmission seasons (bold orange, $95\%$ interval dashed) to be avoided with supplemental vaccine.}
\label{fig:risk_test}
\end{figure*}

Results from this approach are visualized in Fig. \ref{fig:sus_recon}. Estimates via Eq. \ref{eq:reg2} (blue) capture the dynamic range of those from the more mechanistic transmission model (grey). In both cases, sudden drops from vaccination campaigns (pink) and from large outbreaks (black) are balanced by steady increases. While the growth rates in the transmission models vary over time based on changes in routine immunization coverage and birth rates, those based on the seasonality profile alone approximate an appropriate average. Despite some discordance on the effectiveness of particular vaccination campaigns, like in 2016 in Kenya, the estimates are clearly similar.

\section{Intervention timing}

Eqs. \ref{eq:regression}, \ref{eq:stability}, and \ref{eq:reg2} are not a complete transmission model, particularly since $\hat{S}$ remains unspecified, but they still capture key features of the epidemiology. Specifically, in areas with seasonal transmission, we can anticipate when populations need supplemental immunization in order to avoid above average susceptibility during a high transmission season. Or, said differently, we can ask in the wake of a large vaccination effort how long the population can go before needing another \cite{verguet2015controlling}. 

From Fig. \ref{fig:sus_recon}, in the absence of campaigns and large outbreaks, susceptibility grows linearly at rate $\alpha_1 \times I_0/\hat{S}$, and after a large campaign, like in Kenya in 2016 or in Chad in 2021, that linear growth is the dominant effect until susceptibility rebounds to above average values. This observation inspires a coarse forecasting method, namely estimating $Z_t/\hat{S}$ after a large event and then extrapolating linearly.

We test this idea out on a few examples in Fig. \ref{fig:risk_test}. Reconstructions based on partial data (blue) are shown for four countries, simulating decision-making after a campaign effort. For example, in Pakistan entering 2019, we may have wanted to know how the 2018 outbreak and subsequent campaign affected future planning. The linear extrapolation (orange) would have correctly estimated that 2021 was the first upcoming high-season (bold, with dashed lines indicating uncertainty) with 2017-like outbreak risk. Across settings, even with considerable differences in historical campaign and outbreak frequencies, we see similar stories.

While this method is far from a quantitative forecasting platform, one that could for example estimate the size and scale of upcoming measles outbreaks, it's at least a lens with which to view case time series that encapsulates many of the key epidemiological considerations. With such minimalist data needs, just the time series and the record of past campaigns, it gives us an operational tool comparable in applicability to current empirical methods, but with a deeper connection to transmission mechanisms. As an overall approach, it seems to warrant further study.

\section{Conclusion}

The key idea in this paper is that the seasonality profile is a stable epidemiological structure containing a surprising amount of information about transmission. We often think of demographic factors like immunization coverage and birth rates as the drivers of measles transmission processes. In retrospect, it seems reasonable that outbreak structure gives us a view into these same underlying demographic factors. 

As a planning tool, the method encapsulated in Eqs. \ref{eq:regression}, \ref{eq:stability}, and \ref{eq:reg2} have minimal data and complexity requirements while still generating valuable perspective on a population's epidemiological situation. We don't anticipate that they'll replace transmission models and other more sophisticated forecasting tools, but we think they can highlight areas requiring attention at particular times, and their application may also help those more established approaches be more successful.

\bibliography{references}
\end{document}